\documentclass[10pt,amsmath,aps,amssymb,superscriptaddress,notitlepage,onecolumn,longbibliography]{revtex4-1}
\usepackage{graphicx}
\usepackage{float}
\usepackage{graphicx}
\usepackage{amsmath}
\usepackage{amssymb}
\usepackage{bm}
\usepackage{overpic}
\usepackage{anyfontsize}
\usepackage{bigints}
\usepackage{array}
\usepackage{makecell}
\usepackage{stmaryrd}
\usepackage{hyperref}
\usepackage[dvipsnames]{xcolor}
\usepackage{cancel}
\usepackage{sansmath}
\usepackage[normalem]{ulem}

\hypersetup{
	colorlinks,
	citecolor=blue,      
	filecolor=red,
	linkcolor=blue,
	urlcolor=blue,
	hyperfigures
}

\usepackage{todonotes}

\newcommand{\tensbf}[1]{\bm{\mathsf #1}} 

\providecommand\bcdot{\boldsymbol{\cdot}}
\DeclareMathAlphabet{\mathsfbi}{T1}{phv}{b}{it}
\newcommand{\bff}{{\boldsymbol f}}

\newcommand{\bu}{{\boldsymbol u}}

\newcommand{\bx}{{\boldsymbol x}}

\newcommand{\bp}{{\boldsymbol p}}
\newcommand{\md}{{\mathrm{d}}}

\newcommand{\comD}[1]{{\textcolor{black}{#1}}}
\newcommand{\olivia}[1]{\textcolor{black}{#1}}

\begin{document}

\title{Signatures of elastoviscous buckling in the dilute rheology of stiff polymers}

\author{Brato Chakrabarti}
\affiliation{Department of Mechanical and Aerospace Engineering, University of California San Diego, 9500 Gilman Drive, La Jolla, CA 92093, USA}
\affiliation{Center for Computational Biology, Flatiron Institute, New York, NY 10010, USA}
\author{Yanan Liu}
\thanks{Present address: School of Physics, Northwest University, Xi'an, China}
\author{Olivia du Roure}
\author{Anke Lindner}
\affiliation{Laboratoire de Physique et M\'ecanique des Milieux H\'et\`erog\`enes, UMR 7636, ESPCI Paris, PSL Research University, CNRS, Universit\'e \olivia{de Paris}, Sorbonne Universit\'e, 10 rue Vauquelin, 75005 Paris, France}
\author{David Saintillan}
\thanks{Correspondence: \url{dstn@ucsd.edu}}
\affiliation{Department of Mechanical and Aerospace Engineering, University of California San Diego, 9500 Gilman Drive, La Jolla, CA 92093, USA}


\begin{abstract}
As a stiff polymer tumbles in shear flow, it experiences compressive viscous forces that can cause it to buckle and undergo a sequence of morphological transitions with increasing flow strength. We use numerical simulations to uncover the effects of these transitions on the steady shear rheology of a dilute suspension of stiff polymers. Our results agree with classic scalings for Brownian rods in relatively weak flows but depart from them above the buckling threshold. Signatures of elastoviscous buckling include enhanced shear thinning and an increase in the magnitude of normal stress differences. We discuss our findings in the light of past work on rigid \comD{Brownian} rods and non-Brownian \comD{elastic fibers} and highlight the subtle role of thermal fluctuations in triggering instabilities. \vspace{-0.15cm}
\end{abstract}

\maketitle

\section{Introduction}

Understanding and relating bulk rheological properties of complex fluids to the orientations, deformations, and interactions of their microscopic constituents has been a long-standing challenge in fluid mechanics \cite{larson2005rheology,schroeder2018single}.\ With the ability to visualize single molecules using fluorescence microscopy, a large body of research using deoxyribonucleic acid (DNA) has focused over the last two decades on deciphering the dynamics and rheological properties of dilute long-chain polymer solutions in simple flows \cite{shaqfeh2005dynamics}. The persistence length $\ell_p$ of DNA molecules is much shorter than their typical contour length $L$, and in this limit conformational properties are governed by a competition between entropic forces favoring coiled conformations and viscous forces tending to stretch the molecules, as quantified by the Weissenberg number $Wi\equiv\dot{\gamma}\tau_r$, or product of the applied strain rate $\dot{\gamma}$ with the longest polymer relaxation time $\tau_r$. Along with coarse-grained simulations \cite{jendrejack2002} and kinetic theories \cite{winkler2006semiflexible}, these molecular rheology studies have highlighted how microscopic conformational dynamics give rise to macroscopic rheological properties such as shear thinning and normal stress differences in shear  flow \cite{hur2000brownian,schroeder2005dynamics}, and lead to the `coil-stretch' transition in extensional flow \cite{schroeder2003observation}. 

In the other limit, the rheology of rigid Brownian rod-like suspensions is also well understood \cite{doi1988theory}. Thermal fluctuations in this case result in orientational diffusion with rotational diffusivity $d_r = {3 k_B T \ln (2 r)}/{\pi \mu L^3}$, where $k_B T$ is the thermal energy, $\mu$ is the shear viscosity of the solvent, and $r = L/a$ is the aspect ratio of the rods with characteristic radius $a$. The competition between orientational diffusion favoring an isotropic distribution and the background shear that tries to align the rods is characterized by the rotary P\'eclet number $Pe \equiv \dot{\gamma} d_r^{-1}$, where $d_r^{-1}$ describes the orientational relaxation time. With increasing $Pe$, the preferential alignment of these rod-like polymers near the flow axis reduces viscous dissipation, resulting in shear thinning and also yielding nonzero normal stress differences. Three distinct scaling regimes for the shear viscosity and normal stress differences as functions of $Pe$ have been identified and characterized by the foundational theoretical analyses of \citet{leal1971effect}, \citet{hinch1972effect} and \citet{brenner1974rheology} (see \S \ref{sec:rigidrheo} and Table~\ref{tab:tab1} for a summary). However, so far only rigid Brownian rods have been considered in detail, and the role of flow-induced deformations on the rheology of these suspensions remains largely unexplored.

We address this problem here with focus on {stiff} polymers characterized by $\ell_p\gg L$, the opposite limit compared to DNA. While in weak flows these filaments behave as rigid rods, they are known to undergo various buckling instabilities in stronger flows \cite{becker2001instability,manikantan2015buckling,liu2018morphological,duroure2019anrfm,chakra2020helical}, yet clear insight into the specific role of these instabilities in the rheology of {dilute} suspensions is lacking.  Here, we use numerical simulations to relate the morphological transitions of stiff Brownian filaments in simple shear flow to the rheology of their dilute suspensions. We also contrast our predictions with known results for non-Brownian deformable fibers \cite{becker2001instability,tornberg2004simulating} and uncover how they are altered by shape fluctuations and orientational diffusion.

The paper is organized as follows. In \S\,\ref{sec:sec2}, we provide details of the polymer model, measures of the extra stress and a brief summary of the scaling laws for Brownian rigid rods. We present numerical results for the rheology in both two and three dimensions in \S\,\ref{sec:sec3} and discuss our predictions in the context of the rheology of rigid Brownian rods as well as non-Brownian elastic fibers. \vspace{-0.3cm}

\section{Problem description and methodology}\label{sec:sec2}

\subsection{Governing equations}
In the dilute limit, we simulate the dynamics of a single polymer modeled as a fluctuating, inextensible Euler elastica with centerline parameterized as $\bx(s,t)$ where $s$ is arc-length \cite{liu2018morphological}. Hydrodynamics is captured by local slender body theory for Stokes flow, in which the centerline position evolves as \vspace{-0.1cm}
\begin{equation}\label{eq:SBT}
8 \pi \mu \left[\partial_t\bx(s,t) - \bu_\infty\right] = -\tensbf{\Lambda} \bcdot \bff(s,t).  \vspace{-0.1cm}
\end{equation}
Here, $\bu_\infty = (\dot{\gamma} y, 0, 0)$ is the background shear flow with constant shear rate $\dot{\gamma}$. The force per unit length exerted by the filament on the fluid is modeled as $\bff = B \bx_{ssss} - (\sigma \bx_s)_s + \bff^{{b}}$, 
where $B$ is the bending rigidity, $\sigma$ is a Lagrange multiplier that enforces inextensibility of the filament and can be interpreted as line tension, and $\bff^{{b}}$ is the Brownian force density obeying the fluctuation-dissipation theorem. The local mobility operator $\tensbf{\Lambda}$ accounts for drag anisotropy and is given by  \vspace{-0.1cm}
\begin{equation}
\tensbf{\Lambda}\bcdot\bff= \left[(2-c) \tensbf{I} - (c+2) \bx_s \bx_s\right] \bcdot \bff, \vspace{-0.1cm} \label{eq:Lambda}
\end{equation}
where $c = \ln(\epsilon^2 \mathrm{e}) < 0$ is an asymptotic geometric parameter and $\epsilon  = r^{-1}$ is the inverse aspect ratio. This geometrically nonlinear description of the centerline elasticity has been extensively used to successfully describe various elastohydrodynamic problems  in both low \cite{shelley2000stokesian,tornberg2004simulating,young2007stretch,manikantan2013subdiffusive,lim2008dynamics,chakra2020helical} and high Reynolds number flows \cite{allende2018stretching,banaei2020numerical}. The present model is identical to the classical planar Euler elastica problem \cite{singh2019planar,audoly2000elasticity} and is a suitable description as long as the local curvature $\kappa \ll a^{-1}$. In the case of slender filaments used here for which $a\ll L$, such extreme deformations {do not occur} occur over the range of flow strengths considered, thus justifying the use of this model.  Note that the mobility operator of equation (\ref{eq:Lambda}) does not account for non-local hydrodynamic interactions between distant parts of the filaments. Including these interactions can be achieved using the non-local slender-body operator as done in our past work \cite{liu2018morphological}. {This results, however,} in an increased computational cost that is prohibitive for the present study which requires averaging over very long times. To test the consequences of this approximation, we have also performed a few select simulations with full hydrodynamics, where we observed {only} slight quantitative differences with the local drag model in terms of the magnitude of stresses, but no difference in the scalings of the various rheological quantities with respect to flow strength.

We scale lengths by $L$, time by the characteristic relaxation time of bending deformations $\tau_r = 8 \pi \mu L^4/B$, elastic forces by the bending force scale $B/L^2$, and Brownian forces by $\sqrt{\vphantom{A}\smash{L/\ell_p}}\,B/L^2$. The dimensionless equation of motion then reads \vspace{-0.15cm}
\begin{equation}
\partial_t\bx(s,t)= \bar{\mu} \bu_\infty - \tensbf{\Lambda} \cdot \Big[\bx_{ssss} - (\sigma \bx_s)_s + 
\sqrt{\vphantom{A_i}\smash{L/\ell_p}}\,\boldsymbol{\zeta} \Big], \vspace{-0.15cm}
\end{equation}
where $\boldsymbol{\zeta}$ is a Gaussian random vector with zero mean and unit variance. Two dimensionless groups appear: (i) the elastoviscous number \vspace{-0.15cm}
\begin{equation}
\bar{\mu} \equiv \frac{\dot{\gamma}\tau_r}{c}=\frac{8 \pi \mu \dot{\gamma} L^4}{Bc} \vspace{-0.15cm}
\end{equation}
serves as the measure of hydrodynamic forcing against internal elasticity and plays a role analogous to the Weissenberg number for flexible polymers, and (ii) $L/\ell_p$ captures the importance of thermal shape fluctuations. The limit of rigid Brownian rods formally corresponds to $\bar{\mu}\rightarrow 0$ (no flow-induced deformations) and $L/\ell_p\rightarrow 0$ (no thermal shape fluctuations). We note that $\bar{\mu}$ and $L/\ell_p$ are related to the rotary P\'eclet number: \vspace{-0.15cm}
\begin{equation}
Pe = \frac{\bar{\mu} \,c}{24 \ln (2 r)} \frac{\ell_p}{L}, \label{eq:Peclet}  \vspace{-0.15cm}
\end{equation}
which will facilitate comparisons of our simulations of Brownian polymers with finite bending resistance with analytical predictions for rigid Brownian rods. 

\subsection{Numerical methods}

Numerical schemes to solve equation~\eqref{eq:SBT} in the absence of Brownian motion have been described previously in great detail \cite{tornberg2004simulating}, and we provide only a brief outline here. In equation \eqref{eq:SBT}, the line tension $\sigma(s,t)$ which acts as a Lagrange multiplier is an unknown. In order to solve for it, we make use of the inextensibility constraint $\bx_s \cdot \bx_s = 1$. Differentiating this constraint with respect to time and using the slender-body-theory equation provides a second-order linear ordinary differential equation for $\sigma$ as explained in detail by \cite{tornberg2004simulating}, which is subsequently solved with the boundary condition $\sigma = 0$ at $s = 0,1$. The time-marching of equation \eqref{eq:SBT} is performed using an implicit-explicit second-order accurate backward finite difference scheme where the stiff linear terms arising due to bending are treated implicitly while the nonlinear terms are handled explicitly.  The boundary conditions for time-marching are the force- and moment-free conditions for the filament ends, which translate to $\bx_{ss} = \bx_{sss} = 0$ at $s = 0,1$. We used $N=64-128$ points to discretize the filament centerline and typical time steps were in the range of $\Delta t \sim 10^{-6}-10^{-9}$.

Treatment of the spatially and temporally uncorrelated Brownian forces in equation \eqref{eq:SBT} requires special attention and has been described previously by \cite{manikantan2013subdiffusive} and \cite{liu2018morphological}.  Specifically, we apply a low-pass filter to smooth out the noise along the centerline and typically remove 50 percent of the high-frequency components in the process. The algorithm was benchmarked against standard equilibrium properties of semiflexible polymers \cite{wilhelm1996radial}.  \vspace{0.4cm}

\subsection{Measures of stress}
A calculation of the extra stress in a dilute suspension of force- and torque-free particles was provided by \citet{batchelor1970stress}. The single-particle contribution to the bulk stress tensor in the dilute limit is given by the stresslet, which generalizes the Kirkwood formula commonly used for molecular systems \cite{irving1950statistical}. For our polymer model, the expression for the extra stress is \vspace{-0.15cm}
\begin{equation}
\tensbf{\Sigma} =  -n\, \Big\langle\int_{0}^{L} \left[\frac{1}{2}(\bx\bff +\bff\bx)-\frac{1}{3} \mathsfbi{I} (\bx \bcdot\bff)  \right]\md s \Big\rangle, \label{eq:stresslet}  \vspace{-0.15cm}
\end{equation}
where $n$ is the number density in the suspension, $\bff$ is the dimensionless force density exerted  on the fluid with contributions from both elastic deformations and Brownian fluctuations, and $\langle\cdot\rangle$ denotes the ensemble average. This expression is extremely convenient for non-Brownian fibers \cite{becker2001instability}. However, in simulations of Brownian polymers, fluctuations have contributions of $\mathcal{O}(\Delta t^{-1/2})$ where $\Delta t$ is the integration time step. These contributions also enter the Lagrange multiplier $\sigma(s,t)$ that enforces inextensibility, resulting in a poor convergence of the ensemble average as first pointed out by \citet{doyle1997dynamic}. Since our interest is in the steady-state extra stress, we instead use the Giesekus stress expression commonly used for polymers \cite{doyle1997dynamic,ottinger1996stochastic},   \vspace{-0.25cm}
\begin{equation}
\tensbf{\Sigma} =  -n\, \Big\langle\int_{0}^{L} \left[\frac{1}{2}(\bx\,\mathsfbi{R}  \bcdot \bu_\infty +\mathsfbi{R}  \bcdot \bu_\infty\,\bx)-\frac{1}{3} \mathsfbi{I} (\bx \bcdot\mathsfbi{R}  \bcdot \bu_\infty)  \right] \md s \Big\rangle,\label{eq:giesekus}  \vspace{-0.15cm}
\end{equation}
where $\mathsfbi{R} = \tensbf{\Lambda}^{-1}$ is the local resistance tensor along the centerline. Results obtained with equation (\ref{eq:giesekus}) were tested against (\ref{eq:stresslet}) in various regimes of $Pe$. In weak flows ($Pe\ll1$), the Giesekus stress was found to slightly underestimate the magnitude of the viscosity as it omits Brownian contributions, but excellent agreement was found in stronger flows and identical scalings in terms of $Pe$ were obtained by both methods across all regimes of $Pe$. In the following, we present results based on equation (\ref{eq:giesekus}), which is computationally more efficient than the Kirkwood expression.  \vspace{-0.1cm}

\subsection{Summary of rigid rod rheology \label{sec:rigidrheo}}
A slender non-Brownian rod in shear flow undergoes a periodic tumbling motion known as a Jeffery orbit \cite{jeffery1922motion}. During this periodic tumbling, the particle spends most of its time aligned with the flow direction and  equal amounts of time in the extensional and compressional quadrants of the flow.
This dynamics is fundamentally altered in the presence of rotational diffusion. While the shear flow still results in quasi-periodic tumbling, the Brownian rod is now able to stochastically sample different orbits \cite{Zottl2019}, resulting in an anisotropic orientational probability distribution $\psi(\bp)$ at steady state, where $\bp$ is a unit vector that identifies the orientation of the  rod \cite{chen1999orientation}. This distribution leads to a mean orientation of the rod in the extensional quadrant, giving rise to a contractile stresslet as the inextensible rod resists stretching by the flow. This stresslet in turn alters the effective viscosity of the system. In the dilute limit  of $n L^3 \ll 1$, computing the extra stress reduces to obtaining the steady state orientation distribution of a single Brownian rod. Contributions to the stresslet arise from the external flow and from Brownian diffusion and can be computed using slender body theory \cite{batchelor1970stress,leal1971effect,hinch1972effect,brenner1974rheology}:  \vspace{-0.15cm}
\begin{equation}
\tensbf{\Sigma}^{f} = \frac{\pi \mu n L^3}{6 \ln(2 r)} \left[\langle \bp \bp \bp \bp \rangle - \frac{1}{3}\mathsfbi{I}\langle \bp \bp \rangle \right] \boldsymbol{:} \mathsfbi{E}_\infty, \quad	\tensbf{\Sigma}^{b} = 3 n k_B T \left[\langle \bp \bp \rangle - \frac{1}{3}\mathsfbi{I} \right],  \vspace{-0.15cm}
\end{equation}
where $\mathsfbi{E}_{\infty}$ is the rate-of-strain tensor of the applied flow $\bu_\infty$. The  extra stress $\tensbf{\Sigma}=\tensbf{\Sigma}^{f}+\tensbf{\Sigma}^{b}$ is thus entirely determined from the second and fourth moments, $\langle \bp \bp \rangle$ and $\langle \bp \bp \bp \bp \rangle$, of rod orientations. These moments can be computed from the steady state orientation distribution function $\psi(\boldsymbol{p})$, which is set by the balance of the advective rotational flux due to the flow and of the Brownian diffusive flux. \citet{hinch1972effect} solved for the distribution function and associated particle stress in three distinct asymptotic regimes of the rotary P\'eclet number $Pe$. These regimes and corresponding scalings are summarized in Table \ref{tab:tab1}. 
\begin{table}
	\centering
	\renewcommand{\arraystretch}{1.0}\vspace{-0.15cm} 
	\begin{tabular}{cccc}
		\textbf{Regimes}                        & \multicolumn{3}{c}{\textbf{Scalings}}  \\ [-3pt] \hline 
		& $\eta$ & $\Sigma_{xy}$ & $\Sigma_{xx}, \Sigma_{yy}, \Sigma_{zz}$ \\ [3pt] \cline{2-4} 
		\\ [-5pt]
		$Pe \ll 1$                  &  constant & $Pe$ &  $Pe^2$ \\ [4pt]
		$r^3 + r^{-3} \gg Pe \gg 1$ & $Pe^{{-1/3}}$ & $Pe^{2/3}$ & $Pe^{{2/3}}$ \\ [4pt]
		$Pe \gg r^3 + r^{-3}$    &  constant & $Pe$ &  $Pe^2$ \\ \hline                                
	\end{tabular} \vspace{-0.15cm} 
	\caption{ Asymptotic scalings for the relative viscosity $\eta$ and the individual stress components in dilute suspensions of rigid Brownian rods in various regimes of the rotational P\'eclet number $Pe$ and aspect ratio $r$ \cite{hinch1972effect,brenner1974rheology}. The first and second normal stress differences $N_1>0$ and $N_2<0$ scale similarly  as the diagonal stress components.} 
	\label{tab:tab1} 
\end{table}
The three rheological measures of primary interest to us are the relative polymer viscosity $\eta$ and the first and second normal stress differences $N_1$ and $N_2$,  \vspace{-0.1cm} 
\begin{equation}
\eta = \frac{\Sigma_{xy}}{\mu \dot{\gamma} nL^3} , \qquad
N_1 = \frac{\Sigma_{xx}-\Sigma_{yy}}{n k_B T}, \qquad
N_2 = \frac{\Sigma_{yy}-\Sigma_{zz}}{n k_B T}. \vspace{-0.25cm}  \label{eq:viscosity}
\end{equation}

\section{Numerical results and discussion}\label{sec:sec3}	

\begin{figure}
	\centering
	\includegraphics[width=1.0\linewidth]{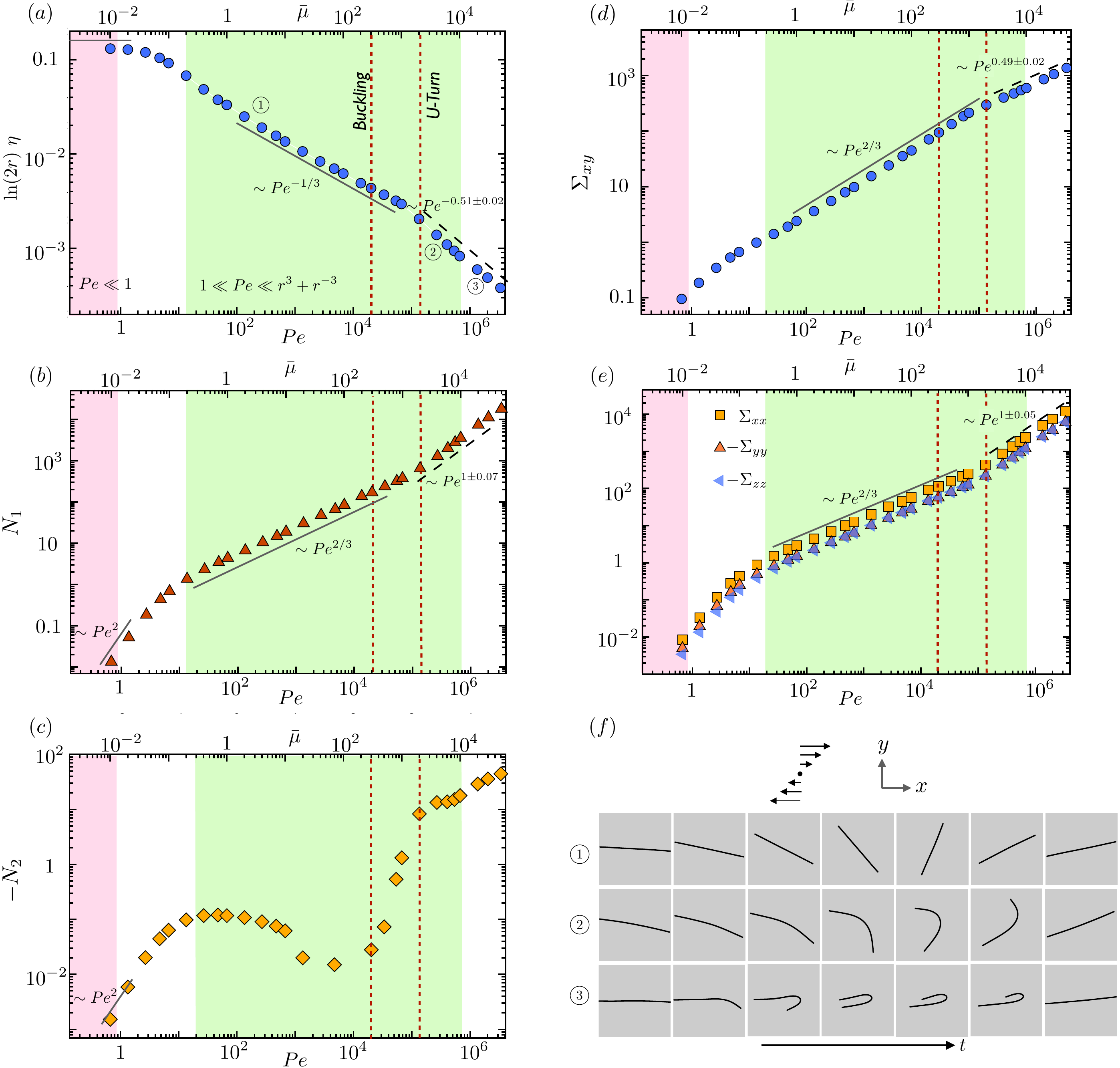}\vspace{-0.0cm}
	\caption{($a$) Polymer viscosity $\eta$, ($b$) first normal stress difference $N_1$, ($c$) negative second normal stress difference $-N_2$, ($d$) shear stress, and ($e$) normal stress components as functions of  $Pe$ (bottom axis) and  $\bar{\mu}$ (top axis) for  polymers with $\ell_p/L=1000$ and $r=220$ in 3D. The low- and intermediate-$Pe$ scalings are theoretical predictions for rigid rods, for which they are valid inside the pink ($Pe\ll1$) and green ($1\ll Pe\ll r^3+r^{-3}$) regions. The vertical dashed lines show the onsets of $C$ buckling and $U$ turns \cite{liu2018morphological}. ($f$) Typical sequences of conformations during tumbling, $C$ buckling, and a $U$ turn. Corresponding regimes and values of $Pe$ are labeled in ($a$). In all figures, marked scalings before the buckling transition are rigid rod predictions (solid line), whereas scalings past the transition are numerical observations (dotted~line).}\vspace{-0.2cm}
	\label{fig:eta3d}
\end{figure}

\subsection{Three-dimensional rheology of stiff polymers}

We present numerical results on the rheology in three dimensions, with focus on the case of stiff slender polymers with $\ell_p/L = 1000$ and aspect ratio $r=220$. In this limit, the main effect of Brownian motion is to cause orientational diffusion with negligible shape fluctuations, and any deformations are thus the result of elastoviscous buckling. Figure~\ref{fig:eta3d}($a,b,c$) shows the relative polymer viscosity $\eta$ and first and second normal stress differences $N_{1,2}$ as functions of P\'eclet number $Pe$ (or, equivalently, elastoviscous number $\bar{\mu}$). In weak flows ($Pe\ll 1$, pink region), $\eta$ exihibits a plateau whereas $N_{1,2}$ both grow from zero as $Pe^2$, with $N_1>0$ and $N_2<0$. With increasing flow strength, shear-thinning takes place as the polymers start aligning with the flow, and a second regime begins with scalings of $\smash{\eta \sim Pe^{-1/3}}$ and $\smash{N_{1} \sim Pe^{2/3}}$ in perfect agreement with the theoretical predictions of Table~\ref{tab:tab1}. The data for the second normal stress difference is very noisy in this range of $Pe$ and fails to capture the expected scaling of ${Pe^{2/3}}$ for reasons we explain below. In this regime, the filament remains straight and tumbles quasi-periodically as shown in the first row of figure~\ref{fig:eta3d}($f$).

As the filament performs a tumble, it rotates across the compressional quadrant of the flow, where it experiences compressive viscous stresses. Above a critical value of the elastoviscous number of $\bar{\mu}^{(1)} = 306.6$, these stresses can overcome bending rigidity and drive an Euler buckling instability leading to deformed configurations reminiscent of a $C$ shape, typical of the first mode of buckling \cite{becker2001instability}. The filament then rotates as a $C$ and stretches out again as it enters and sweeps through the extensional quadrant. With increasing flow strength, the filament becomes more likely to buckle while remaining nearly aligned with the flow direction, and this ultimately gives rise to distinctive folded $U$ shaped conformations that perform tank-treading motions while maintaining a mean orientation close to the flow axis \cite{harasim2013direct}. As uncovered in our past work \cite{liu2018morphological}, the transition to this new mode of transport occurs at $\bar{\mu}^{(2)} = 1.8 \times 10^3$. Both of these thresholds are indicated by vertical lines in figure~\ref{fig:eta3d}($a,b,c$) and, for the chosen values of $\ell_p/L$ and $r$, fall within the intermediate scaling range of $1\ll Pe \ll r^3+r^{-3}$ (green region). Typical conformations from {tumbling}, $C$ buckling and $U$ turns are  shown in figure~\ref{fig:eta3d}($f$).

Quite remarkably, we find that the onset of buckling has no immediate signature on the rheology, with the Brownian rod-like scaling laws persisting past the critical value of $\bar{\mu}^{(1)}$. This result is in contrast with the two-dimensional rheology of non-Brownian elastic fibers studied by \cite{becker2001instability} and \cite{tornberg2004simulating}, where buckling is responsible for shear thinning and nonzero normal stress differences. This discrepancy is attributed to the presence of 3D rotational diffusion in our simulations. In shear flow, the viscous compressive force experienced by the filament is a function of its orientation and reaches a maximum at an angle of $3 \pi/4$ with the direction of flow in the plane of shear. In the presence of rotational noise, the filament orientation is not restricted to the shear plane and the  maximum compression experienced is reduced. This translates to a set of measure zero for the probability density function $\psi(\boldsymbol{p})$, and therefore the probability of a buckling event is negligible at $\bar{\mu}=\bar{\mu}^{(1)}$. As 
$\bar{\mu}$ is increased beyond $\bar{\mu}^{(1)}$, buckling becomes increasingly more likely, and indeed $\eta$ and $N_{1,2}$ start to depart from the intermediate scalings before the onset of tank-treading in figure \ref{fig:eta3d}. This departure marks a transition to new  scalings of $\smash{\eta \sim {Pe}^{{-0.51\pm 0.02}}}$ and $N_1 \sim Pe^{{1 \pm 0.07}}$ and is accompanied by a sharp increase in $N_2$. These rheological changes are clear signatures of elastoviscous buckling, as they occur within the range of validity of the intermediate rigid rod scalings. 

A more complete picture is provided in figure~\ref{fig:eta3d}($d$,$e$), showing the shear and diagonal components of the extra stress tensor. In particular, we find that normal stresses in figure~\ref{fig:eta3d}($b$) are dominated by $\Sigma_{xx}$, while $\Sigma_{yy}$ and $\Sigma_{zz}$, which are negative, have smaller magnitudes. All three components follow the same scaling, with the intermediate scaling of $Pe^{2/3}$ giving way to a nearly linear scaling into the buckling regime. Note that for $Pe\gg 1$ the values of  $\Sigma_{yy}$ and $\Sigma_{zz}$ are almost identical, which explains the strong noise in the data for $N_2$ in figure~\ref{fig:eta3d}($c$), especially in the intermediate scaling regime.

\begin{figure}
	\centering
	\includegraphics[width=1\linewidth]{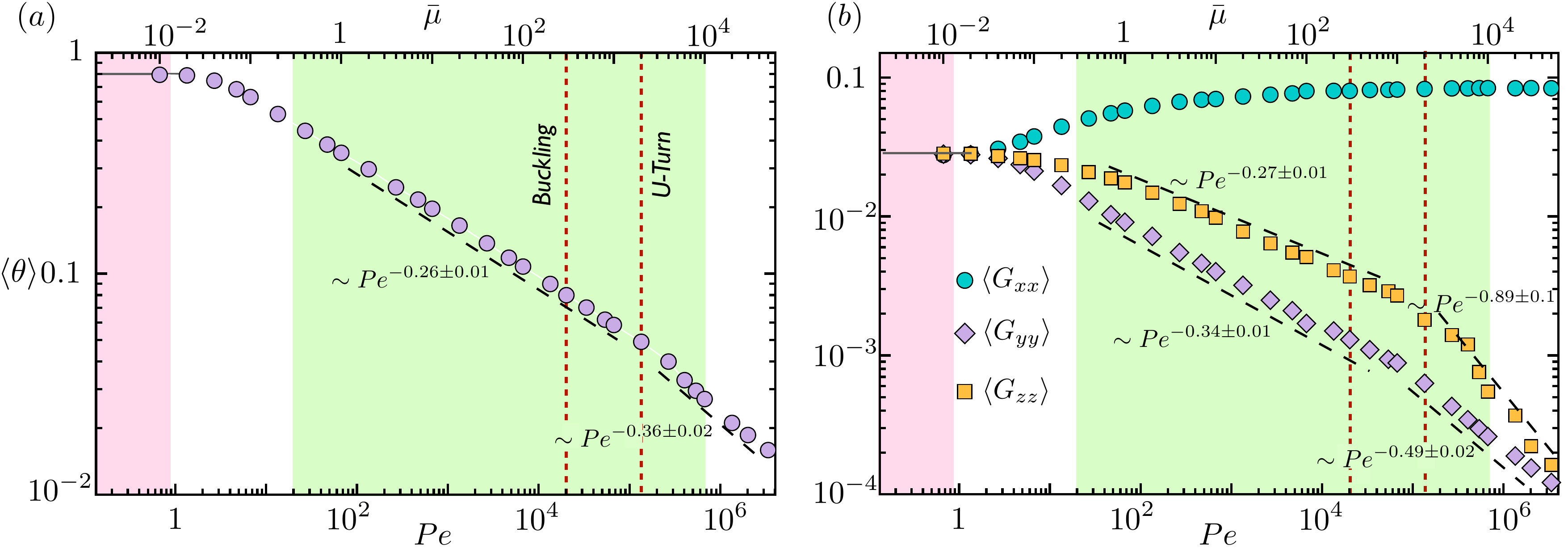}\vspace{-0.0cm}
	\caption{($a$) Mean polymer orientation $\langle\theta\rangle$, defined as the angle made by the dominant eigenvector of the gyration tensor with the flow direction, as a function of $Pe$ and $\bar{\mu}$. ($b$) Diagonal components $\langle G_{xx}\rangle$, $\langle G_{yy}\rangle $ and $\langle G_{zz}\rangle$ of the mean gyration tensor.   } \vspace{-0.2cm}
	\label{fig:gyr}
\end{figure}

To relate these findings to filament conformations, we introduce the gyration tensor \vspace{-0.1cm}
\begin{equation}
\mathsfbi{G}(t) = \int_0^1 \left(\bx - \bx_c \right) \left(\bx - \bx_c \right) \mathrm{d}s, \vspace{-0.1cm}
\end{equation}
whose dominant eigenvector is used to define the mean filament orientation $\theta(t)$ with respect to the flow direction. Its ensemble average is shown as a function of P\'eclet number in figure \ref{fig:gyr}($a$). In weak flows ($Pe\ll 1$), $\langle \theta \rangle$ asymptotes to the value of $\pi/4$ for an isotropic orientation distribution, and correspondingly the diagonal components of $\langle\mathsfbi{G}\rangle$, which describe the variance of the polymer mass distribution along the coordinate directions, all tend to the same value of $1/4\pi$ in figure \ref{fig:gyr}($b$). Alignment of the filament with the flow is accompanied by a decrease in the mean orientation angle $\langle\theta\rangle$ with increasing shear rate, and is also indicated by the growth and saturation of $\langle{G}_{xx}\rangle$ while $\langle G_{yy}\rangle$ rapidly decays. Increasing flow strength also forces the filament toward the shear plane leading to a decrease in $\langle{G}_{zz}\rangle$ as well. In strong flows, the initiation of  $U$ turns leads to a sharper decrease in both $\langle G_{yy}\rangle$ and $\langle G_{zz}\rangle$, as the emergent folded conformations remain increasingly aligned with the flow direction as seen in the third row of figure \ref{fig:eta3d}($f$). 

It is interesting to note that the scaling law for the viscosity is altered at a slightly lower value of $\bar{\mu}$ compared to the gyration tensor or mean orientation angle. This solidifies the idea that the signature of deformations observed in the stress and viscosity stems from the gradually more frequent occurrence of buckling events. During $C$ buckling (second row of figure~\ref{fig:eta3d}($f$)), the deformed filament still tumbles as a whole, leading to negligible changes in the behavior of the gyration tensor and mean orientation angle, even though the stress components are affected.    \vspace{-0.2cm}

\subsection{Discussion}

{Above, we have discussed the case of stiff polymers and have compared our results to known scaling laws for rigid rods in three dimensions. Stiff polymers and rods both experience strong rotational diffusion. Shape fluctuations are absent in the case of rigid rods and remain weak for the stiff polymers in our simulations performed in the limit of $\ell_p/L\gg 1$. The main difference between the two systems is thus the occurrence of buckling instabilities above a given threshold for the stiff polymers.} Numerical results show a clear signature of such elastoviscous buckling on the rheology, with enhanced shear thinning and normal stress differences compared to the case of rigid rods.

We now address the role of shape fluctuations on the rheology. Remaining in the limit of weak fluctuations for stiff polymers, we can tune fluctuations by varying $\ell_p/L$ while keeping rotational diffusion important, with a smaller $\ell_p/L$ corresponding to stronger shape fluctuations. For simplicity, we present here results obtained in 2D, and show in 
figure~\ref{fig:lp} the variation of the shear viscosity $\eta$ and normal stress difference $N_1$ as functions of shear rate for {two combinations of ($\ell_p/L$, $r$). Note that the theoretical threshold for buckling is identical in 2D and 3D, since the first instance of buckling occurs for a filament lying in the shear plane. The theoretical onset of $U$ turns was predicted in our past work using a 2D reduced-order model \cite{liu2018morphological}, but also faithfully describes the transition in 3D since the dynamics is primarily two-dimensional in strong flows as previously shown in figure~\ref{fig:gyr}($b$).
	Scaling laws have been determined for both $\eta$ and $N_1$ before and after the buckling threshold, with a transition region where the scaling is changing continuously. 
	Before the threshold, the measured scalings are in agreement with 3D rigid rod predictions within error bars, and quantitatively identical results are obtained for both values of $\ell_p/L$ since elasticity plays a negligible role in that regime. 
	As the P\'eclet number increases, buckling occurs first for the smaller value of  $\ell_p/L$, since $Pe$ and $\ell_p/L$ are related by {equation} (\ref{eq:Peclet}) and the buckling threshold occurs at a fixed $\bar{\mu}=\bar{\mu}^{(1)}$. Beyond the threshold, close agreement is found between 2D and 3D results, and identical scalings are obtained for $\eta$ and $N_1$ for both values of $\ell_p/L$, indicating that varying the importance of shape fluctuations, while remaining in the limit of weak fluctuations, does not alter the observed rheology noticeably.
	The limit of strong shape fluctuations, relevant to a number of experimental systems and to {previous} work \cite{liu2018morphological, harasim2013direct}, is outside the theoretical and numerical framework developed here.

	\begin{figure}[H]
		\centering
		\includegraphics[width=1\linewidth]{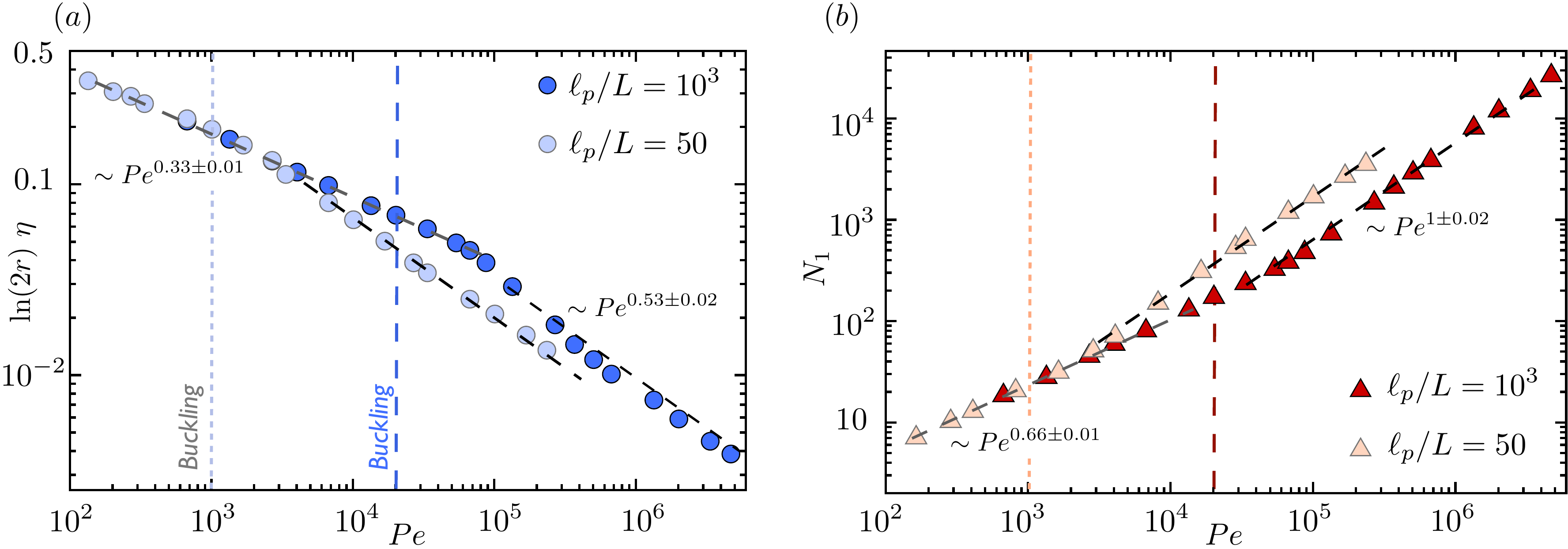}\vspace{-0.15cm}
		\caption{($a$) Polymer viscosity $\eta$, and ($b$) normal stress difference $N_1$ as functions of $Pe$ for $(\ell_p/L,r)=(1000,220)$ and $(50,100)$ in two-dimensional simulations. Vertical dashed lines show the two buckling thresholds, which correspond to the same value of $\smash{\bar{\mu}=\bar{\mu}^{(1)}}$.} \vspace{-0.25cm}
		\label{fig:lp}
	\end{figure}
	
	Another observation can be made from figure~\ref{fig:lp}. In three dimensions in figure~\ref{fig:eta3d}, the changes in the scaling laws of viscosity and normal stress differences do not occur right at the theoretical onset of buckling, but are delayed due to the presence of rotational diffusion, as the probability for the polymer to align perfectly with the direction of maximum compression is negligible right at the buckling threshold.  On the contrary, in 2D, any polymer performing a tumble is required to sweep through the entire compressional quadrant and is therefore more likely to buckle. This interpretation is confirmed by our 2D resuits in figure~\ref{fig:lp}, in which the change in scaling due to deformations now occurs slightly closer to the theoretical  buckling threshold. Nevertheless, and in contrast to fully non-Brownian systems \cite{becker2001instability}, the transition is not abrupt, likely due to the presence of Brownian noise. 
	The fact that the slope changes occur closer to the buckling threshold confirms again that the occurrence of $C$ buckling, rather than $U$ turns, is responsible for the change in behavior. In addition, we find that the Pence of $U$ turns is not accompanied by another clear change in scaling. Rather, both buckling events and $U$ turns have the same signature on the rheology, in spite of their distinct morphologies and dynamics.

	Our results can also be discussed in comparison to previous observations made for non-Brownian elastic fibers in 2D. For non-Brownian fibers, shape fluctuations as well as rotational diffusion are absent. \citet{tornberg2004simulating} and \cite{becker2001instability} predicted the onset of normal stress differences and shear thinning above the buckling threshold from 2D simulations.  In contrast to the case of Brownian rods, normal stress differences are zero and the shear viscosity is constant in the absence of buckling. No scaling laws for $\eta$ {and $N_1$} exist for non-Brownian fibers in the dilute limit. 
	
	{Simulations performed with non-Brownian fibers require the use of an initial perturbation for the buckling instability to be triggered, and the properties chosen might influence the obtained results. Random fluctuations are naturally present in Brownian systems as investigated here and thus no initial perturbation needs to be imposed. To connect our Brownian simulations more directly to the non-Brownian case, we perform a numerical experiment, focusing for the sake of illustration on a regime where $\bar{\mu} > \bar{\mu}^{(2)}$ and where the dynamics is dominated by $U$ turns with occasional $S$ shaped modes \cite{liu2018morphological}. In this experiment, we perform a standard Brownian simulation with $\ell_p/L=50$ but artificially switch off  Brownian forces $\boldsymbol{f}^b$ at the initiation of any buckling event, detected by the threshold of $R_{ee}/L < 0.98$ on the end-to-end distance. In this way, we produce initial conditions set by Brownian noise and thus identical to those present in our Brownian simulations, but can investigate the arising stresses for a situation without Brownian noise. Stresses are only evaluated once the noise has been switched off. This artifical situation does not allow a meaningful viscosity to be calculated and thus we show $\Sigma_{xy}$ and $N_1$ in figure \ref{fig:fictstress}.
		These stresses are plotted vs $\bar{\mu}$, which is more relevant than $Pe$ in the case of non-Brownian fibers that do not experience orientational diffusion. Recall that $\bar{\mu}$ and $Pe$ are proportional to each other in equation (\ref{eq:Peclet}), while  $\eta$ and $\Sigma_{xy}$ are related through $\dot{\gamma}$ in equation (\ref{eq:viscosity}). 
		
			\begin{figure}[H]
			\centering
			\includegraphics[width=1\linewidth]{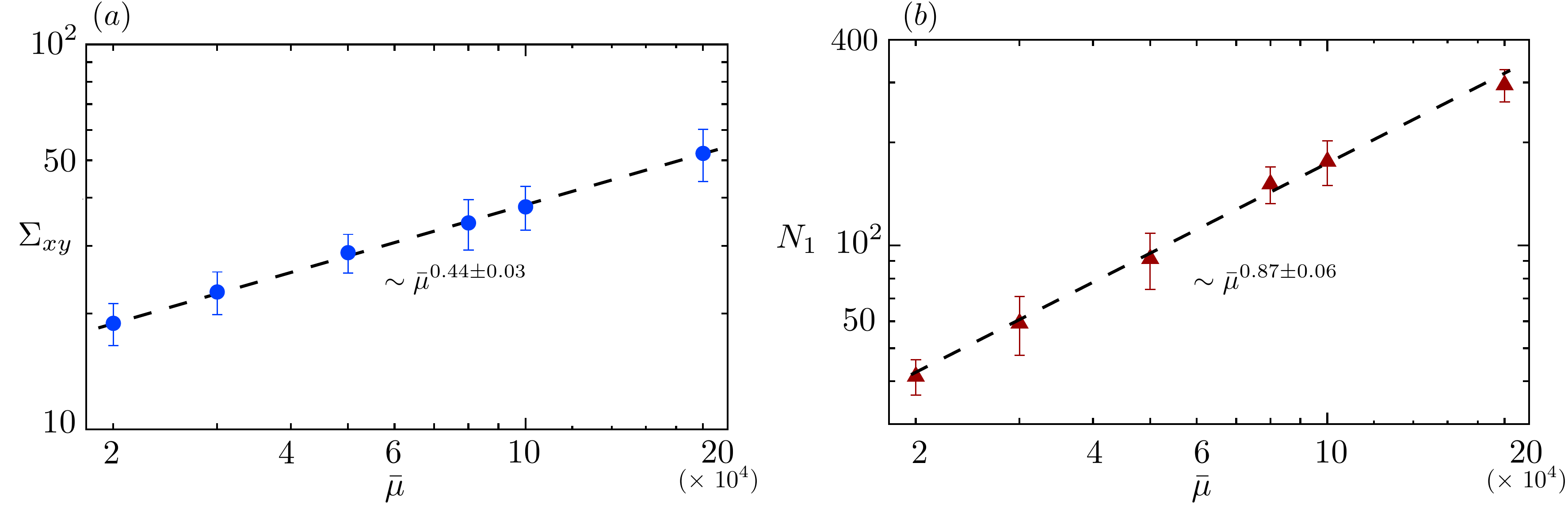}  \vspace{-0.2cm}
			\caption{Numerical experiment illustrating the respective roles of elastic instabilities and thermal fluctuations. Perturbations for these simulations are obtained from a Brownian filament with $\ell_p/L=50$. Variation of ($a$) shear stress and ($b$) first normal stress difference vs $\bar{\mu}$.}
			\label{fig:fictstress}  \vspace{-0.3cm}
		\end{figure}
		
		The  scalings obtained for $\Sigma_{xy}$ and $N_1$ with respect to $\bar{\mu}$ translate into $\eta  \sim \Sigma_{xy}/\dot{\gamma} \sim Pe^{-0.56 \pm0.03}$ and $N_1 \sim Pe^{0.87 \pm0.06}$. Surprisingly, these exponents differ only slightly from those observed in the fully Brownian simulations in figure~\ref{fig:lp}}.
	As this numerical experiment does not account for the Brownian stress that arises due to shape fluctuations and orientational diffusion during buckling, our results suggest that the leading contribution to the stress and its scaling with flow strength is set by elastic instabilities, with fluctuations mainly serving to trigger polymer tumbles as in the rigid rod case while also exciting the dominant buckling modes. 
	Note that \cite{becker2001instability} found $N_1<0$ for non-Brownian fibers in sufficiently strong flows, in disagreement with our findings. We believe this discrepancy may be due to the particular way the fiber backbones were perturbed in their simulations.

Our numerical simulations have cast light on the role of elastohydrodynamic instabilities on the rheology of dilute suspensions of {stiff} polymers in the limit of $\ell_p\gg L$. The leading effect of buckling was shown to enhance shear-thinning in strong flows while also driving an increase in normal stress differences in comparison to the case of rigid Brownian rods. Detailed rheological measurements in the dilute regime and in monodisperse systems are a challenge and have yet to  be performed, but would be of great use to confirm our numerical predictions and connect them with past observations in more concentrated systems \cite{huber2014,kirchenbuechler2014,lang2019,miglani2018thermal,devarajan2020thermal}. Extensions of the present work may consider the case of semi-flexible polymers with $L\sim\ell_p$, which is more challenging numerically, as well as the role of hydrodynamic interactions in semi-dilute suspensions.

\vspace{-0.2cm}

\begin{acknowledgments}
	The authors thank M. Shelley for useful discussions. A.L., B.C. and Y.L. acknowledge funding from the ERC Consolidator Grant PaDyFlow (Agreement 682367).\ D.S. acknowledges funding from a Paris Sciences Chair from ESPCI Paris. 
\end{acknowledgments}

\bibliography{bibfile}

\end{document}